\documentclass[prl,twocolumn,superscriptaddress]{revtex4}

\usepackage{multirow}
\usepackage{graphicx}
\usepackage{amsmath}
\usepackage{enumerate}
\usepackage{epstopdf}
\usepackage{times}
\usepackage{color}
\usepackage{bm}
\usepackage{comment}
\usepackage{braket}
\usepackage{accents}
\usepackage{subfig}

\begin{document}

\title{
Entanglement in ground and excited states of gapped fermion systems and their relationship with fermi surface and thermodynamic equilibrium properties
}

\author{Michelle Storms}
\affiliation{Department of Physics, University of California Davis, CA 95616, USA}
\affiliation{Department of Physics, Ohio Wesleyan University, Delaware, OH 43015, USA}
\author{Rajiv R. P. Singh}
\affiliation{Department of Physics, University of California Davis, CA 95616, USA}

\date{\rm\today}

\begin{abstract}
We study bipartite entanglement entropies in the ground and excited states of model fermion systems, where a staggered potential, $\mu_s$, induces a gap in 
the spectrum. 
Ground state entanglement entropies satisfy the `area law', and the `area-law'
coefficient is found to diverge as a logarithm of the staggered potential,
when the system has an extended Fermi surface at $\mu_s=0$. On the square-lattice, we show that 
the coefficient of the
logarithmic divergence depends on the fermi surface geometry and its orientation
with respect to the real-space interface between subsystems and is
related to the Widom conjecture as enunciated by Gioev and Klich (Phys. Rev. Lett. 96, 100503 (2006)). 
For point Fermi surfaces in two-dimension, the `area-law' coefficient stays finite as $\mu_s\to 0$.
The von Neumann entanglement entropy associated with the excited states follows a `volume law' and allows us to 
calculate an entropy density function
 $s_{V}(e)$, which is substantially different from the thermodynamic entropy density function $s_{T}(e)$, 
when the lattice is bipartitioned into two equal subsystems but 
approaches the thermodynamic entropy density as the
fraction of sites in the larger subsystem, that is integrated out, approaches unity.

\end{abstract}


\maketitle

\section{Introduction}

In recent years, the study of quantum entanglement properties of many-body systems has been a very active
area of research.
Bipartite entanglement in the ground state of quantum many body systems is
known to satisy an `area law' (with at most logarithmic corrections) \cite{cardy,rmp_review,cramer}, 
i.e., when a large system is divided into two subsystems,
the entanglement entropy between them is proportional to the `area' measuring the
boundary between subsystems \cite{hastings}. Such an `area law' has been shown to be at the heart of the success of
density matrix renormalization group (DMRG) and tensor network based variational methods for
computing ground state properties of many-body model Hamiltonians \cite{dmrg,tensor-network}.
While the Hilbert space of a many-body
system grows exponentially with the `volume' of the system, the computational requirements of these
methods grow exponentially only with the `area' of the subsystems and possibly only polynomially
in the size of the system when full potential of tensor network methods is realized. So far, this has already allowed high
precision calculations of properties of large many body systems in one-dimension and on 
finite diameter cylinders \cite{kagome}.

Another important role of the study of entanglement properties has been in identifying new phases
and in characterizing continuous phase transitions. The `area law' is known to be logarithmically
violated for gapless one-dimensional systems, and this violation is well understood within the framework
of conformal field theory \cite{cardy}. In higher dimensions,
singularities often associated with sub-leading terms in the entanglement entropy, can provide universal
signatures of critical phenomena without reference to specific order parameters \cite{vidal},
while long-range entanglement helps identify topological phases
and spontaneously broken symmetries \cite{levin_wen,kitaev,isakov,kagome2}. Methods to calculate entanglement properties of 
interacting many-body systems in $d>1$ are being actively explored \cite{melko-swap,nlc,grover,grover2}.

Another subject that has been a topic of much recent research activity is that of
`thermalization' \cite{deutsch,srednicki}. Especially motivated by many beautiful experiments in
cold atom systems, several researchers have explored the issues of long time
behavior of isolated quantum systems prepared far from equilibrium \cite{rigol,rigol2,cassidy,cazalilla,calabrese,essler}. 
One set of questions relate to a quench, where a system starts out at some instance
in a pure quantum state that is far from an eigenstate of the system. Are the local properties in
such a state, after a long time, described by a thermal density 
matrix? It was
found that integrability of the model plays an important role in answering this question.
Integrable systems, with large number of constants of motion, may not
`thermalize' in the conventional sense. Rather than being described by the
usual thermal density matrix, their long time behavior is described by
more generalized statistical ensembles \cite{rigol2}.
Also, calculations of von Neumann entanglement
entropy after a quench was found to have clear deviations from the thermal entropy \cite{peschel}.

Here, we study the entanglement entropy of gapped fermion systems by the
correlation-matrix method \cite{peschel,casini,lehur}, as well as the series expansion method \cite{book}. We consider
a half-filled system of free spinless fermions with hopping paramater $t$. We add
a staggered chemical potential, $\mu_s$, which introduces a gap in the spectrum. 
Series expansions are developed for the second Renyi entropy in powers of $t/\mu_s$.
We find an excellent agreement between the series expansion and correlation-matrix
methods, where both calculations have been performed. The importance of the series expansion
calculation is that it can be readily applied to interacting fermi systems and may provide
an elegant way to study changes in fermi surfaces and onset of gaps in such systems.

As expected, we find that the ground state entanglement entropy of these gapped systems satisfies an `area-law'.
However, the area-law term is logarithmically divergent in the parameter $t/\mu_s$, as $\mu_s\to 0$,
as long as the system has an extended fermi surface at $\mu_s=0$. We study variations in this area-law
term by introducing an anisotropy in the hopping matrix elements. We find that
the coefficient of the logarithmic divergence is not universal but varies with anisotropy.
We use a straightforward generalization of the Widom conjecture as enunciated by Gioev and Klich \cite{klich,wolf,haas,dmrg-1d}
to relate this variation
with anisotropy to the shape of the fermi surface and its relative orientation with respect to the interface
between subsystems. For systems with point fermi-surfaces in two-dimension, such as with $\pi$-flux fermions, 
we find that there is no logarithmic divergence. Instead, the area-law coefficient remains finite
as $\mu_s\to 0$.

We also calculate the bipartite entanglement entropy 
associated with the excited states. In this case,
we randomly select an ensemble of states with different energy per site $e$. We calculate
the bipartite von Neumann entanglement entropy associated with the excited states and find that
it is proportional to the volume of the system and
can be used to define an entropy density function $s_{v}(e)$, which is sharply defined
for large systems. Comparison of this quantity 
with the conventional thermal entropy density
function $s_{T}(e)$ shows intersting behavior. When the system is subdivided into two equal parts (A and B),
and part B is integrated out, the von Neumann entanglement entropy associated with A is always much
smaller than the thermal entropy. However, if we divide the system into two unequal parts A and B,
and integrate out the degrees of freedom of the larger part B, the von Neumann entanglement entropy density
of part A approaches the thermal entropy density when the number of sites in B becomes much larger than
the number of sites in A. This implies that despite this being an integrable system, if we consider
a typical eigenstate of the system and integrate out a large fraction of the system, the remaining
subsystem has a canonical entropy density corresponding to the energy 
available to the system \cite{deutsch13,santos}.
In other words, the larger subsystem simply acts as a heat bath for the smaller subsystem.
This result is a clear numerical demonstration of the `strong typicality' hypothesis  of Santos {\it et al} \cite{santos}, 
who had argued that this result should hold regardless of integrability.

The plan of the paper is as follows. In Section II, we discuss our basic models and methods.
Section III provides a discussion of series expansions and extrapolations. In Section IV, we present results
for the ground state entanglement. In section V, we present results for the excited state
entanglement properties. Finally, in Section VI, we present our conclusions.

\section{Basic Definitions and Methods}

Throughout this work, we will consider a lattice of atomic orbitals, which are divided into two
disjoint subsystems A and B.
The reduced density matrix for subsystem A is obtained from the full density matrix of the system by 
tracing out the degrees of freedom associated with those in its complement B:
\begin{equation}
\rho_A =  \underaccent{B}{\textrm{Tr}}(\rho) = \underaccent{B}{\textrm{Tr}} (\ket{\Psi}\bra{\Psi}).
\end{equation}
The von Neumann entanglement entropy is defined as
\begin{equation}
S = -\textrm{Tr}(\rho_A \ln{\rho_A}).
\label{vndef}
\end{equation}
The Renyi entropy with index $\alpha$ is defined as
\begin{equation}
S_\alpha = \frac{1}{1-\alpha} \ln[\textrm{Tr}(\rho_A^\alpha)].
\label{rdef}
\end{equation}
It is well known that, in the limit $\alpha\to 1$, the Renyi entropy reduces to the von Neumann entropy.

We study the spinless free-fermion Hamiltonian 
\begin{equation}
{\cal H}=-\sum_{i,j} t_{i,j} (\ c_i^\dagger c_j + c_j^\dagger c_i \ ) +\mu_s \sum_i (-1)^i n_i.
\label{hamiltonian}
\end{equation}
The first term, representing fermion hopping, is a sum over all nearest neighbor bonds of a bipartite lattice.
The term $(-1)^i$ is $+1$ on one sublattice and $-1$ on the other. With the uniform chemical potential set
to zero, the system is half filled by particle-hole symmetry. We will consider the cases of dimensionality
$d=1$, $2$ and $3$, but mostly focus on $d=2$ square-lattice. In 2d, we will, in general, consider the hopping
matrix element along x axis $t_x$ to be different from the hopping matrix element $t_y$ along the y axis.
We denote $t_y=t$ and $t_x= qt$. Since an overall energy scale does not change the states of the system,
we compute entanglement properties as a function of $\mu_s/t$ and $q$. In addition, we will also consider
a $\pi$-flux case, where the product of hopping matrix elements in every elementary plaquette is negative \cite{affleck}.

The Hamiltonian in Eq.~\ref{hamiltonian} is diagonalized in momentum space.  The diagonalized Hamiltonian is of the form
\begin{equation}
{\cal H} = \sum_{\vec{k}} [\ \epsilon_-(\vec{k}) \alpha_k^\dagger \alpha_k + \epsilon_+(\vec{k}) \beta_k^\dagger \beta_k \ ],
\end{equation}
where the sum is over the reduced Brillouin zone.
Due to the staggered potential term, momentum is conserved modulo the antiferromagnetic wavevector. In the reduced
Brillouin zone, there are two states for each wavevector $\vec{k}$. The lower energy state has negative energy
$\epsilon_-(\vec{k})$ and the higher energy state has positive energy $\epsilon_+(\vec{k})$. 
In the half-filled ground state, $\left< \alpha_k^\dagger \alpha_k \right> = 1$ and $\left<\beta_k^\dagger \beta_k \right> = 0$.  

To study entropies associated with excited states, we randomly sample the eigenstates 
of the system keeping the filling the same.
One way to do so is to first randomly select a wavevector, then randomly choose whether to 
place an electron in the lower or upper energy state at that wavevector by setting  
either $\left< \alpha_k^\dagger \alpha_k \right>$ or $\left<\beta_k^\dagger \beta_k \right> = 1$
and the other to zero.  
However, this does not produce the most general excited state. In the most general case, one can
pick any half of the available single-particle states to be occupied. 
We found that excited state entropies, obtained in either case were essentially the same and depend only on the energy
per particle of the excited state.

\begin{figure*}[htb]
\begin{center}
  \subfloat{\label{fig:arealaw:1}\includegraphics[width=55mm,angle=-90]{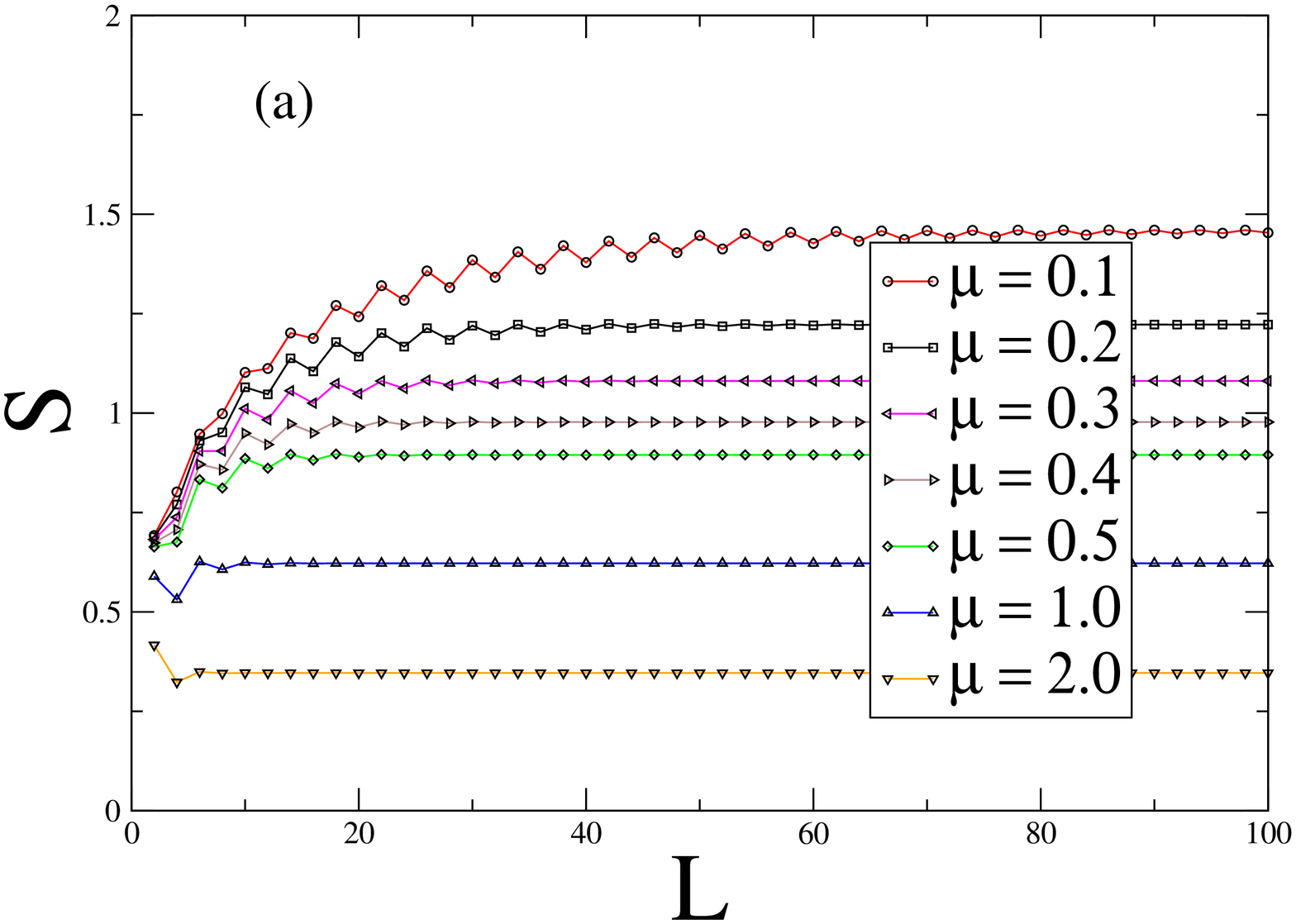}}
  \subfloat{\label{fig:arealaw:2}\includegraphics[width=55mm,angle=-90]{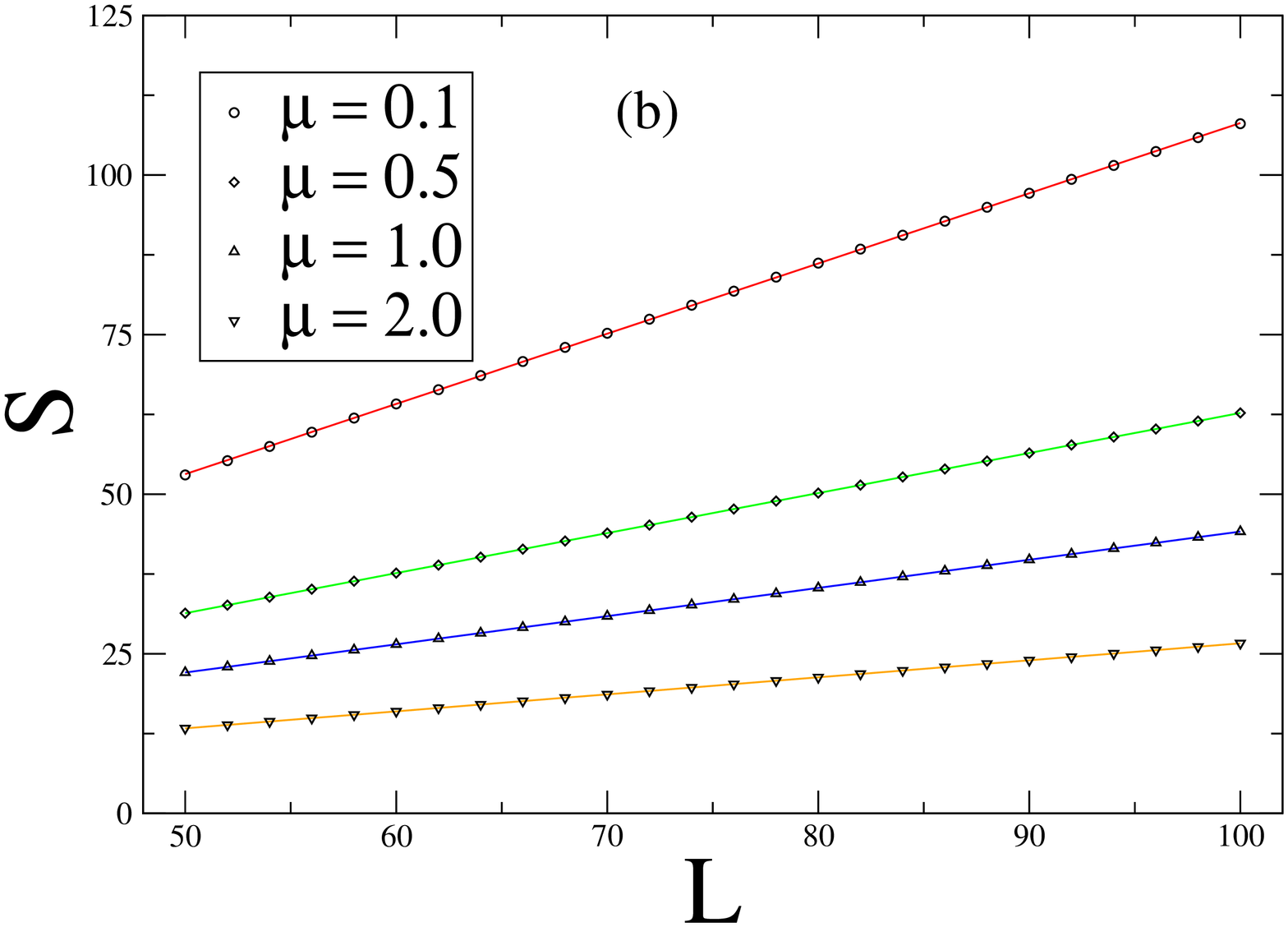}}\\
  \subfloat{\label{fig:arealaw:3}\includegraphics[width=55mm,angle=-90]{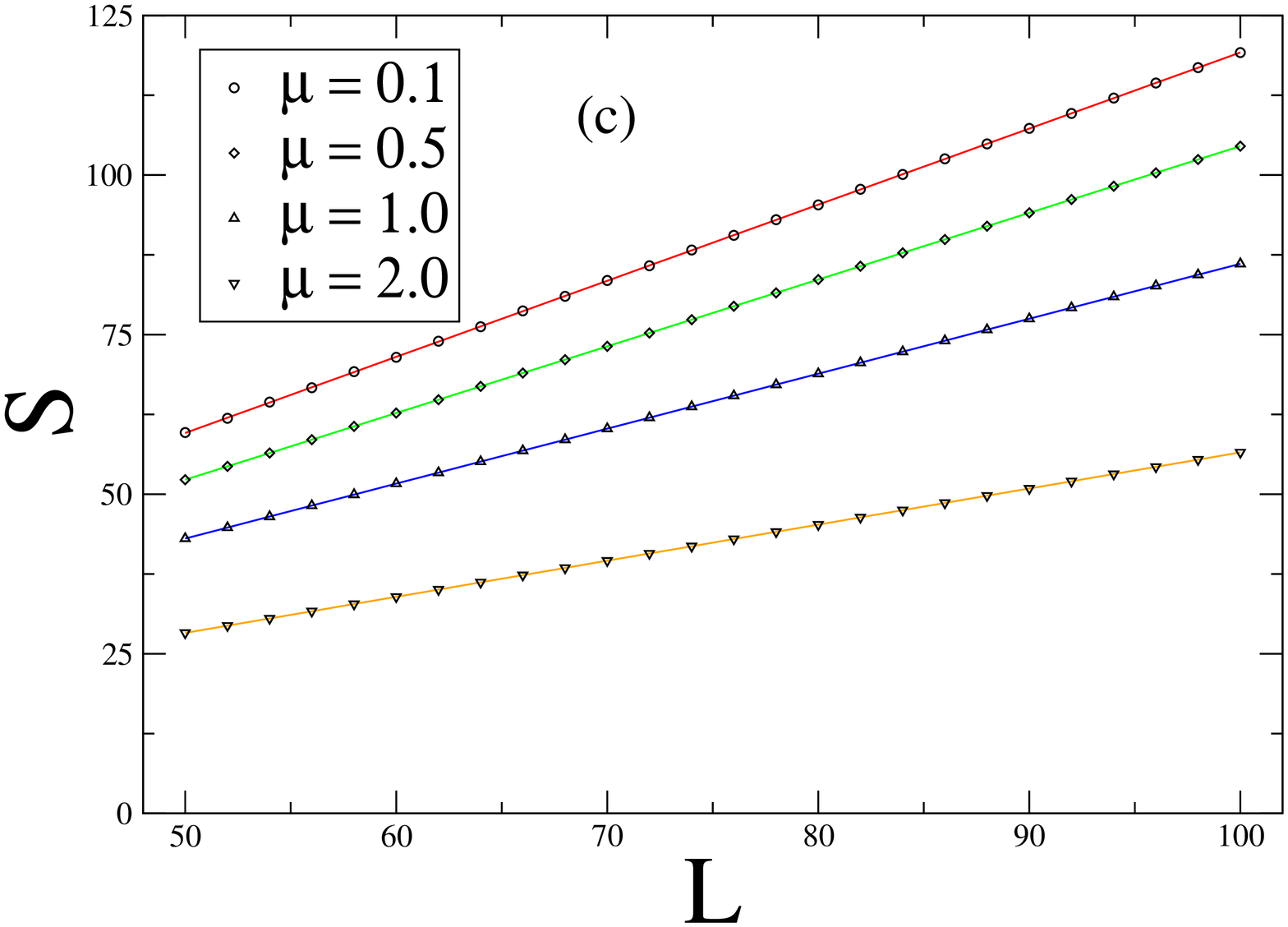}}
  \subfloat{\label{fig:arealaw:4}\includegraphics[width=55mm,angle=-90]{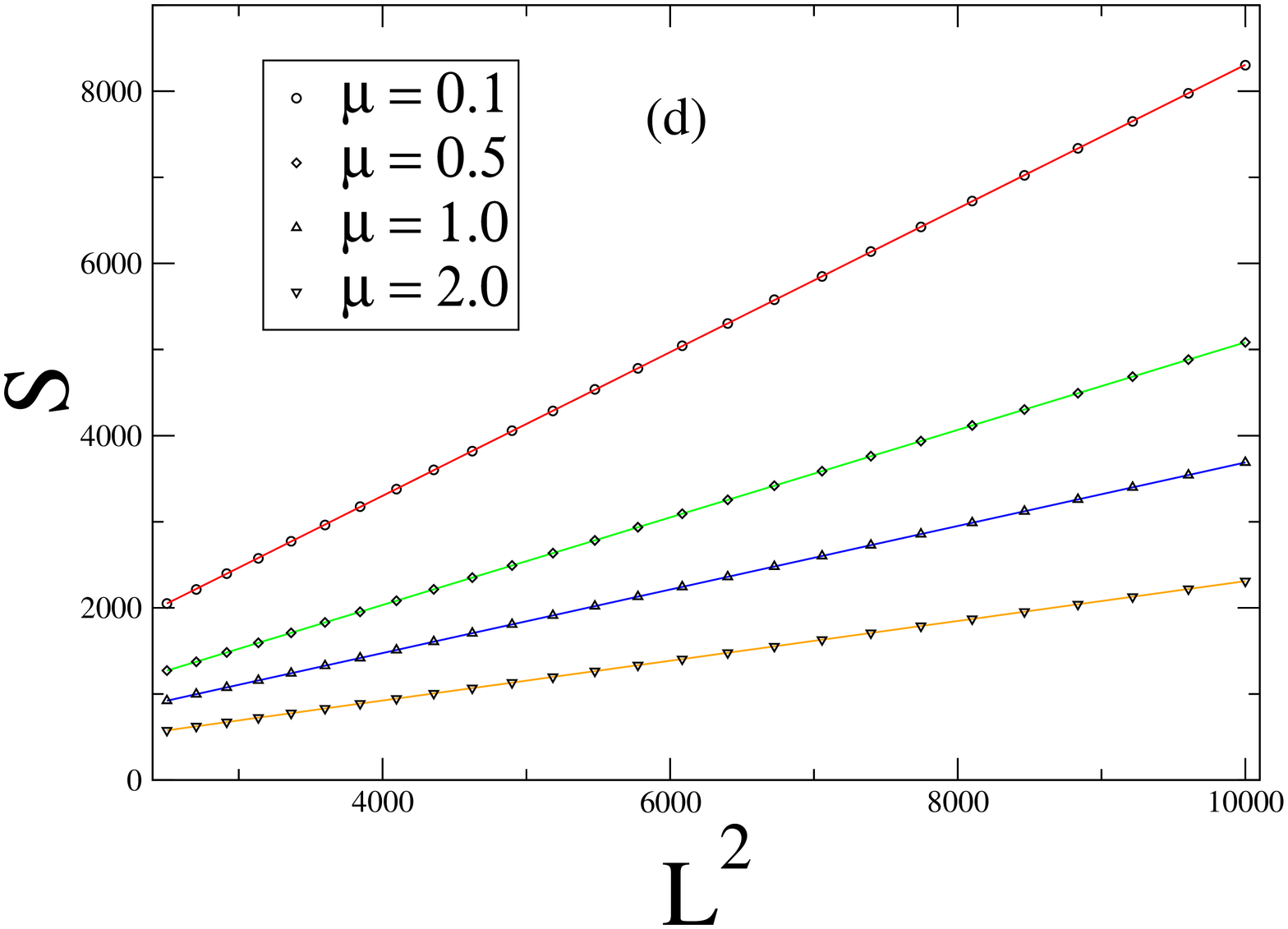}}\\
\caption{von Neumann entanglement entropy in the ground state for different systems with linear size $L$ for
(a) $d=1$, (b) $d=2$ uniform hopping model, (c) $d=2$ $\pi$-flux model and (d) $d=3$ model. The area-law
is valid in all cases.}
\label{fig:arealaw}
\end{center}
\end{figure*}

For non-interacting models, the entanglement entropies can be calculated
in terms of the eigenvalues of the correlation matrix,\cite{peschel,lehur} defined to be
\begin{equation}
{\cal C}_{i,j} = \left< c_i^\dagger c_j \right>,
\end{equation} 
where $i$ and $j$ are restricted to sites in A.
The matrix is readily obtained by expressing it in terms of $\alpha_k$, $\alpha_k^\dagger$, $\beta_k$, and $\beta_k^\dagger$. 
It is diagonalized numerically.
Once the eigenvalues $\lambda_i$ have been obtained, the von Neumann entropy can be expressed as
\begin{equation}
S = - \sum\limits_i \lambda_i \ln(\lambda_i) + (1-\lambda_i)\ln (1-\lambda_i).
\label{vncor}
\end{equation}
Also, the Renyi entropies become,
\begin{equation}
S_\alpha = \frac{1}{1-\alpha}\sum\limits_i \ln[\lambda_i^\alpha + (1-\lambda_i)^\alpha].
\label{rcor}
\end{equation}

Note that the size of the correlation matrix grows as $L^d$ for a system of linear size $L$ in $d$ spatial dimensions, 
making even two-dimensional systems hard to diagonalize for large system sizes.  However, 
if the lattice is split into A and B only along one spatial direction, x, y, or z, 
then the momentum in the perpendicular directions
is conserved and the correlation matrix becomes block diagonal, with size of each block scaling as $L$.  
Each block can then be diagonalized separately and the block eigenvalues can be used directly 
to calculate the entanglement entropy, giving us a much more efficient method \cite{hastings-comm}.
Most of the results we present are for systems with linear dimension $L\le 100$. However, in some cases,
we have studied systems upto $800\times 800$, which still required only very
modest computational resources (calculations for several parameter sets could be completed on a single personal computer in a day).

\section{Series expansions}
Renyi entanglement entropies can be calculated in a power series in the variable $t/\mu_s$ by the
linked cluster method \cite{series-prl,oitmaa}. In this method, one can imagine dividing
an infinite square-lattice into two halves by a straight line running parallel to one of the axes,
and series expansions can be developed for the entanglement entropy per unit length of the boundary
between the subsystems. We have calculated the first few expansion coefficients for the
second Renyi entropy of the zero-flux and the $\pi$-flux hopping models of spinless fermions.
The series for the zero-flux model to order $x^{14}$, with $x=t/\mu_s$, is
\begin{eqnarray}
s_2&=0.5 x^2 -3.125 x^4 + 29.1666667 x^6 -319.3125 x^8 \nonumber \\
&+3805.1 x^{10} -47805.1667 x^{12} + 622826.571 x^{14} 
\end{eqnarray}
while the corresponding series for the $\pi$-flux model is
\begin{eqnarray}
s_2&= 0.5x^2 -2.125 x^4  +10.6666667 x^6 -59.0625 x^8 \nonumber \\
&+349.6 x^{10} -2171.58333 x^{12} +13984.5714x^{14} 
\end{eqnarray}

The series are convergent for small $t/\mu_s$, but to study the singularity at large $x$
we first transform the series to a variable $y=x/(1+x)$. In the new variable the singularity 
is at $y=1$. To build in a logarithmic singularity, when appropriate, we
first take a derivative of the series and then obtain a Pade approximant, which is biased to
have a simple pole at $y=1$. Integrating the Pade approximant builds in a logarithmic singularity at $y=1$.

For the $\pi$-flux case, such a biased extrapolation shows very poor convergence, implying
the absence of a logarithmic singularity. Instead, in this case, we expect the results at
large $t/\mu_s$ to vary with the absolute value of $\mu_s/t$. Thus, after changing to
a variable $y=x/(1+x)$, we further change variables to
\begin{equation}
\delta=1-\sqrt{1-y},
\end{equation}
and then calculate Pade approximants in the variable $\delta$. 

After the analysis,
all results can be transformed back to the original variable $t/\mu_s$.
The results of the series analysis
are presented in the next section together with other numerical results.

\section{Results for Ground State Entanglement}

\begin{figure}[htb]
\includegraphics[scale=.3,angle=-90]{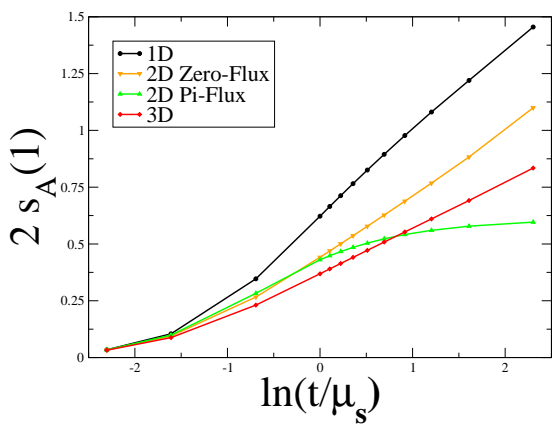}
\caption{Coefficient of the `area-law' term associated with the von Neumann entropy
in different dimensions, including the $\pi$-flux case
in 2d.}
\label{fig:lineterms}
\end{figure}

In this section, we discuss the ground state entanglement entropy for different parameters.
We consider an $L^d$ system (with even $L$), which is divided into two equal parts along the $x$-axis.
The ground-state entanglement is calculated using the methods discussed in Section II.
Fig. \ref{fig:arealaw} shows the observed area-law behavior for 
the von Neumann entropy in d=1, 2, and 3. In two-dimensions,
we have considered both the zero-flux and $\pi$-flux cases.  
In the one-dimensional case, the entropy was found to strongly oscillate with $L$ for small sizes but then
settles down to a constant value, independnt of $L$ at large $L$. In the two and three-dimensional cases, the
entropy scales as $L^{d-1}$ clearly showing the `area law ' behavior. We define the `area-law' coefficient
from the slope
\begin{equation}
S_\alpha=2 s_A(\alpha) \ L^{d-1} +C.
\end{equation}
Note that the boundary between subsystems $A$ and $B$ has two parts, each with area $L^{d-1}$.
Hence the slope of $S_\alpha$ versus $L^{d-1}$ is defined as $2 s_A(\alpha)$.
In Fig.~\ref{fig:lineterms}, we show how the `area-law' coefficient behaves as a function of $t/\mu_s$. 
We find that, except for the $\pi$-flux case, in all other cases, the area-law coefficient 
scales as $\ln{t/\mu_s}$ as $\mu_s\to 0$. We should note that in the $\pi$-flux case, the
fermi surface at $\mu_s=0$ consists of points and has co-dimension two relative to the Brillouin zone.
In this case, the area-law coefficient saturates to a finite value. The logarithmic divergence
is tantamount to $\ln{\xi}$ behavior as the correlation length of the system diverges 
as an inverse power of the staggered potential $\mu_s$ as one approaches the gapless limit ($\mu_s=0$).
It is equivalent to the well known $\ln{L}$ behavior in the gapless systems.\cite{cardy,rmp_review,max}

\begin{figure}[ht]
\centering
\includegraphics[width=0.8\columnwidth,angle=270]{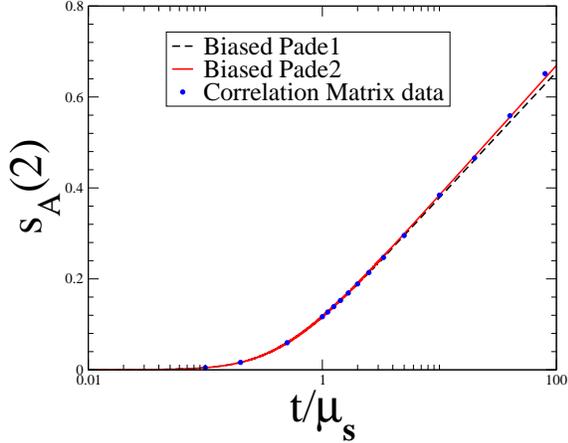}
\caption{Area law coefficient for second Renyi entropy for spinless fermions in two dimensions.
Series expansion results are compared with results from Correlation Matrix method.}
\label{fig2}
\end{figure}

\begin{figure}[ht]
\centering
\includegraphics[width=0.8\columnwidth,angle=270]{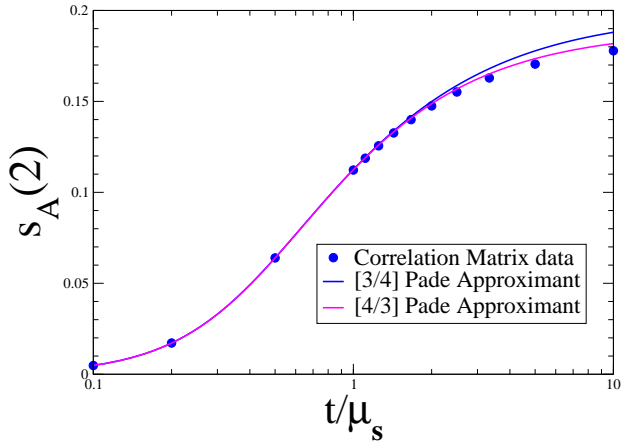}
\caption{Area law coefficient for second Renyi entropy for spinless fermions with pi flux in two dimensions.
Series expansion results are compared with Correlation Matrix method.}
\label{fig3}
\end{figure}

In Fig.~\ref{fig2} and \ref{fig3}, we show comparisons of the second-Renyi entropy area-law coefficient
obtained from series expansion and the correlation matrix methods. The agreement is excellent.
The series expansion is exact at small $t/\mu_s$. But biased series extrapolation also allows us to
accurately capture the singular behavior at large $t/\mu_s$. This is important as
this method can be easily applied to interacting many-fermion systems without any increase in
computational requirements. This could be a potentially powerful numerical method to study changes
in fermi surfaces and the onset of gaps in such interacting systems.

\begin{figure}[ht]
\centering
\includegraphics[width=0.8\columnwidth,angle=270]{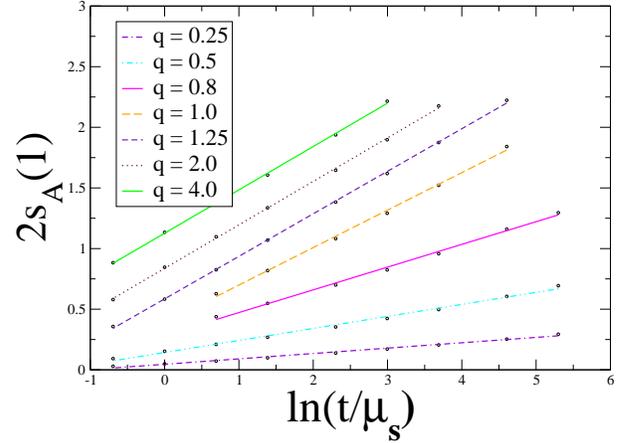}
\caption{von Neumann entanglement entropy with different values of anisotropy $q=t_x/t_y$ on the square-lattice.}
\label{fig5}
\end{figure}

\begin{figure}[ht]
\centering
\includegraphics[scale=.3,angle=-90]{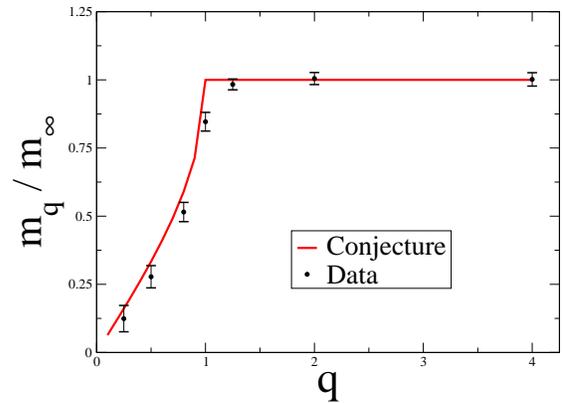}
\caption{Coefficient of the log-divergence in the
area-law coefficient normalized to the 1D case and compared with the Widom conjecture.}
\label{fig:widom}
\end{figure}

In the two-dimensional zero-flux case, the `area-law' coefficient for different anisotropy parameter $q$
is shown in Fig.~\ref{fig5}.  It is clear that the logarithmic divergence for large $t/\mu_s$ obtains for all $q$
values. However, the coefficient of the logarithmic divergence varies with $q$. We call the
asymptotic slope in the $s_A(1)$ versus $\ln{t/\mu_s}$ graph $m(q)$.
We expect this slope to be related to the coefficient of the $\ln{L}$ divergence in the gapless systems.
The latter is known to be governed by the Widom Conjecture as enunciated by  Gioev and Klich \cite{klich}.
In the limit $q\to \infty$, our system simply reduces to decoupled chains and hence in this case,
the divergence is simply equal to that in the one-dimensional case as there is one chain
per unit length \cite{swingle,sidel}.
The Widom Conjecture implies that the slope should vary as 
\begin{equation}
m(q) \propto \int_{FS}(\hat{n} \cdot \hat{x}) ds,
\end{equation}
where the integral is over the fermi surface (FS) at $\mu_s=0$,
$\hat{n}$ is the unit vector normal to the fermi surface and $\hat{x}$ is the unit vector normal to our physical boundary
(in our case this normal to the boundary is fixed to be along the x-axis).  
The coefficient of proportionality can be determined from the one-dimensional limit obtained by letting $q$
go to infinity. The comparison, with no further adjustable parameters is shown in Fig.~\ref{fig:widom}.
Once again, the agreement is very good and confirms the scaling of the area-law coefficient with the
geometry of the fermi surface.

\section{Results for Excited State Entanglement}
\begin{figure}[ht]
\includegraphics[scale = .3, angle = -90]{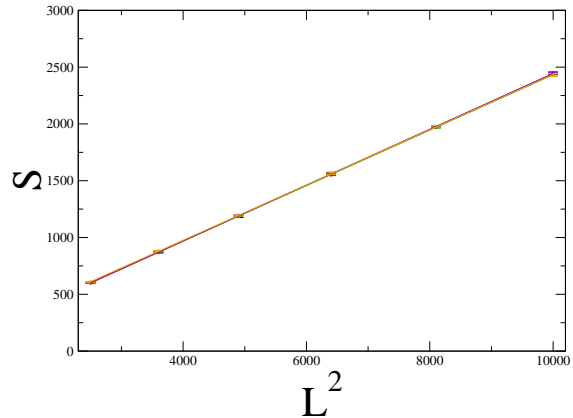}
\caption{Entanglement entropy for the excited states in the 2-dimensional system scales with the
volume of the system.}
\label{fig:bulk}
\end{figure}

In this section, we discuss the results for the entanglement entropy associated with the excited states.
The procedure for sampling the excited states was discussed earlier.
In this case,
we expect the entanglement entropy to follow a "volume law". Indeed, the plots in Fig. \ref{fig:bulk}
confirm this behavior. For the excited states, we have studied a range of $\mu_s/t$ values.
The "volume law" applies regardless of the parameters of the model. 

The entanglement entropy density should depend on the energy density of the eigenstate selected. Let us
define the entanglement entropy density function $s_V(e)=S/V_A$ as a function of the energy density $e=E/V$,
where $V_A=N_A$ is the number of sites in subsystem A, and $E$ and $V=N$ are the total energy and
total volume of the system.
We also vary the number of sites in
A ($N_A$) relative to B ($N - N_A$).
We divide the allowed energy into 
several small intervals. In each interval we generate a number of eigenstates of the system
and calculate the entanglement entropy density. In all cases, we find that entropy density function
is sharply defined with very small variation from eigenstate to eigenstate.
Furthermore, this entropy density has very little
size dependence, showing that our results reflect the thermodynamic limit.

These results can be compared with the grand-canonical 
thermal entropy density for a given energy density $e$, which can be
defined as
\begin{equation}
s_T(e) =-< f(\epsilon) \ln{f(\epsilon)}+ 
(1-f(\epsilon))\ln{(1-f(\epsilon))}>,
\end{equation}
where the angular brackets represent an average over the single-particle states of energy $\epsilon$
and the expression is to be normalized
to give the entropy per site, 
the temperature $T$ is determined by demanding that the energy density at that temperature be $e$,
 and
$f(\epsilon)$ is the fermi distribution function at temperature $T$ 
\begin{equation}
f(\epsilon) = \frac{1}{e^{\epsilon/T} + 1}.
\end{equation}

\begin{figure}[ht]
\centering
\includegraphics[width=0.8\columnwidth,angle=270]{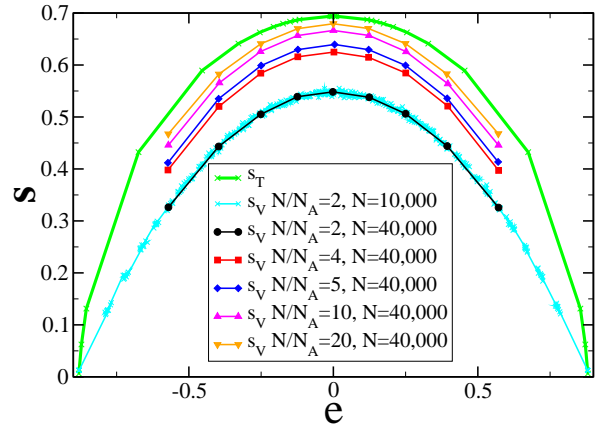}
\caption{Entanglement entropy density of the excited states compared with the thermal entropy density.
The results shown are for $\mu_s/t=0.5$.}
\label{fig6}
\end{figure}

\begin{figure}[ht]
\centering
\includegraphics[width=0.8\columnwidth,angle=270]{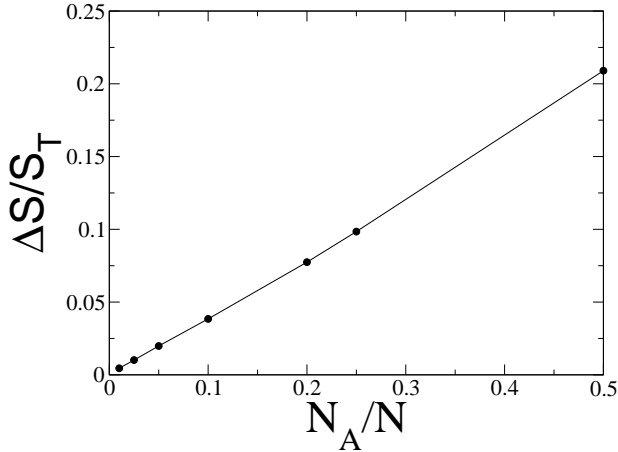}
\caption{Difference between thermal and von Neumann entanglement entropies at $e=0$ as
a function of $N_A/N$. The dark circles represent the calculated data points. The line is 
a guide to the eye.}
\label{fig7}
\end{figure}

In Fig.~\ref{fig6}, we show comparisons of the Von Neumann entanglement entropy density with
the thermal entropy density. Results are shown for two different system sizes ($N=100\times 100$
and $N=200\times 200$). For, the smaller system size ($N=10,000$) results for all the individual
eigenstates sampled are shown, whereas  for the larger size ($N=40,000$), we have
averaged the results for a number of states that are close in energy. In that case, the error bars are negligible
compared to the symbols shown. Also, shown is the entropy density function as the fraction of sites
in subsystem $A$ is reduced compared with the total number of sites. We see that
the entanglement entropy density is substantially lower than the thermal entropy density
when $N/N_A=2$, however it gradually approaches the thermal entropy density when $N/N_A$ becomes large.
Fig.~\ref{fig7} shows the deviation of the thermal entropy from the entanglement entropy normalized
to the thermal entropy for different $N_A/N$ values at $e=0$. It is clear that only as $N_A/N$ goes to zero do the two
entropies become equal. Our results show that as $N_A/N$ goes to zero, the entire von Neumann entanglement
entropy density function approaches the thermal entropy density.

Strictly speaking, it is only when $N>>N_A$ that one should expect the much larger subsystem B to act as
a heat bath for the smaller system A. This equality of the two entropies has been called `strong typicality' by
Santos {\it et al} \cite{santos}, who also defined $f$ as the fraction of sites
that are integrated out. In our notation $f= 1- N_A/N$. The expectation from their
work is that the von Neumann entanglement
entropy should equal thermal entropy at some value of $f$ between one-half and one. For our model, 
the equality only applies when $f$ approaches unity. Whether this is because of the integrability
of the model studied, or is true for all systems, remains to be answered. For non-integrable systems, numerical
studies are mostly going to be limited to relatively small system sizes \cite{santos} due to the exponentially
growing size of the Hilbert space. Then, surface effects are going to be non-negligible,
and may make it difficult to answer this question in
a definitive manner.

\section{Conclusions}

In conclusion, in this paper we have studied the bipartite entanglement entropies of free fermion systems,
where a staggered chemical potential, $\mu_s$, is introduced to induce a gap in the spectrum.
A number of different models have been studied. They include studies of linear chain, square-lattice
and simple cubic lattice with uniform hopping parameter $t$. In the square-lattice case, we have also considered the case of
anisotropic hopping parameters with $t_x\ne t_y$, as well as the $\pi$-flux case.

In the ground state, all these gapped systems obey the `area law', that is, the entanglement entropy
scales with the area measuring the boundary between subsystems. However, as the staggered potential $\mu_s\to 0$, the
area-law coefficient becomes singular. With the exception of the $\pi$-flux phase, in all other
cases we find a $\ln{t/\mu_s}$ divergence as $\mu_s$ goes to zero. This is the analog of the
well-known $\ln{L}$ behavior in the gapless fermi systems. In the $\pi$-flux case, the Fermi-surface
has co-dimension two relative to the Brillouin zone, and such a logarithmic divergence is absent.
We find that the coefficient of the logarithmic divergence follows the Widom Conjecture. 

Comparison of the results for series expansion in $t/\mu_s$ and the correlation matrix method shows excellent
agreement, including the singular behavior at small $\mu_s$. The series expansion method can be
readily applied to interacting models. This could be a useful method to monitor changes in
fermi surface and onset of gaps and phase transitions in interacting fermi systems from a purely
ground state based calculation. It would be interesting to also apply such calculations to
interacting bose systems to look for extended bose-surface singularities \cite{fisher} via
the calculation of entanglement properties.

The entanglement entropies of the excited states show a `volume law' and lead to a well defined
entropy density, which depends on the energy density of the excited state. 
The von Neumann entanglement entropy density is substantially less than the thermal entropy density,
when the system is divided into two equal parts but approaches the thermal entropy density as
the fraction of the system integrated out approaches unity. This provides a definitive demonstration of
`strong typicality' hypothesis put forward by Santos {\it et al} \cite{santos}.

\begin{acknowledgements}

We thank Matt Hastings, Marcos Rigol and Tarun Grover for may useful suggestions and for
a careful reading of the manuscript.
This work is supported in part by NSF grant number  DMR-1004231
and REU grant number PHY-1263201.
\end{acknowledgements}

\end{document}